\documentclass[aps,superscriptaddress,preprintnumbers,showpacs]{revtex4}
\usepackage{amssymb}
\usepackage{graphicx}
\usepackage{dcolumn}
\usepackage{bm}

\begin{document}

\newcommand*{\sjtu}{INPAC, Department of Physics, Shanghai Jiao Tong University, Shanghai, China}\affiliation{\sjtu}
\newcommand*{\NTU}{CTS, CASTS, and Department of Physics, National Taiwan University, Taipei, Taiwan}\affiliation{\NTU}
\newcommand*{\NCTS}{Department of Physics, National Tsing Hua University, and National Center for Theoretical Sciences, Hsinchu 300, Taiwan}\affiliation{\NCTS}

\title{Comment on Reparametrization Invariance \\of Quark-Lepton Complementarity}

\author{Guan-Nan Li}\affiliation{\sjtu}
\author{Hsiu-Hsien Lin}\affiliation{\NTU}
\author{Xiao-Gang He}\email{hexg@phys.ntu.edu.tw}\affiliation{\sjtu}\affiliation{\NTU}\affiliation{\NCTS}

\begin{abstract}
We study the complementarity between quark and lepton mixing angles (QLC), the sum of an angle in quark mixing and the corresponding angle in
lepton mixing is $\pi/4$. Experimentally in the standard PDG parametrization, two such relations exist approximately.
These QLC relations are accidental which only manifest themselves in the
PDG parametrization. We propose reparametrization
invariant expressions for the complementarity relations
in terms of the magnitude of the elements in the
quark and lepton mixing matrices. In the exact QLC limit, it is found that
$|V_{us}/V_{ud}| +  |V_{e2}/V_{e1}| + |V_{us}/V_{ud}| |V_{e2}/V_{e1}| =1$
and $|V_{cb}/V_{tb}| + |V_{\mu 3}/V_{\tau 3}| +|V_{cb}/V_{tb}|| {V_{\mu 3}}/V_{\tau 3}| =1$.
Expressions with deviations from exact complementarity are  obtained.
Implications of these relations are also discussed.

\end{abstract}


\maketitle

Mixing between different generations of fermions in weak interaction
is one of the most interesting issues in particle physics. The mixing is
described by an unitary matrix in the charged current
interaction of W-boson in the mass eigen-state of fermions. Quark
mixing is described by the
Cabibbo~\cite{cabibbo}-Kobayashi-Maskawa~\cite{km}(CKM) matrix
$V_{\rm{CKM}}$, and lepton mixing is described by the
Pontecorvo~\cite{pontecorvo}-Maki-Nakawaga-Sakata~\cite{mns} (PMNS)
matrix $V_{\rm{PMNS}}$ with
\begin{eqnarray}
L = -{g\over \sqrt{2}} \overline{U}_L \gamma^\mu V_{\rm CKM} D_L
W^+_\mu - {g\over \sqrt{2}} \overline{E}_L \gamma^\mu V_{\rm PMNS}
N_L W^-_\mu + H.C.\;,
\end{eqnarray}
where $U_L = (u_L,c_L,t_L,...)^T$, $D_L = (d_L,s_L,b_L,...)^T$, $E_L
= (e_L,\mu_L,\tau_L,...)^T$, and $N_L = (\nu_1,\nu_2,\nu_3,...)^T$
are the left-handed fermion generations. For n-generations, $V =
V_{\rm CKM}$ or $V_{\rm PMNS}$ is an $n\times n$ unitary matrix.

A commonly used form of mixing matrix for three generations of fermions is given by~\cite{ck,FERMILAB-PUB-10-665-PPD},
\begin{eqnarray}
V_{PDG} = \left(
\begin{array}{ccc}
c_{12}c_{13} & s_{12}c_{13} & s_{13}e^{-i\delta}           \\
-s_{12}c_{23}-c_{12}s_{23}s_{13}e^{i\delta} &
c_{12}c_{23}-s_{12}s_{23}s_{13}e^{i\delta}  & s_{23}c_{13} \\
s_{12}s_{23}-c_{12}c_{23}s_{13}e^{i\delta}  &
-c_{12}s_{23}-s_{12}c_{23}s_{13}e^{i\delta} & c_{23}c_{13}
\end{array}
\right),\label{fp}
\end{eqnarray}
where $s_{ij}=\sin\theta_{ij}$ and $c_{ij}=\cos\theta_{ij}$ are the
mixing angles and $\delta$ is the CP violating phase. This is the so called standard PDG parametrization.
If neutrinos
are of Majorana type, for the PMNS matrix one should include an
additional diagonal matrix with two Majorana phases ${\rm
diag}(e^{i\alpha_1/2},e^{i\alpha_2/2},1)$ multiplied to the matrix
from right in the above. The two CP violating Majorana phases do not
affect neutrino oscillations. To distinguish different CP violating phases, the phase
$\delta$ is sometimes called Dirac CP violating phase.
In our later discussions, we will indicate the mixing angles with superscriptions
$Q$ and $L$ for quark and lepton sectors respectively
when specification is needed.

There are a lot of experimental data on the mixing patterns in both
the quark and lepton sectors. For quark mixing, the ranges of the
magnitudes of the CKM matrix elements have been very well determined
with~\cite{FERMILAB-PUB-10-665-PPD}
\begin{eqnarray} \left(
  \begin{array}{lll}
    0.97428\pm0.00015             & 0.2253\pm0.0007    &0.00347^{+0.00016}_{-0.00012}              \\
    0.2252\pm0.0007              & 0.97345^{+0.00015}_{-0.00016}  & 0.0410^{+0.0011}_{-0.0007}      \\
    0.00862^{+0.00026}_{-0.00020} & 0.0403^{+0.0011}_{-0.0007}    & 0.999152^{+0.000030}_{-0.000045}
  \end{array} \right)\;.\label{vv}
\end{eqnarray}
From the above, we obtain the ranges for mixing angles $\theta^Q_{ij}$,
\begin{eqnarray}
\theta^Q_{12}=13.021^\circ\pm0.039^\circ,\quad
\theta^Q_{23}=2.350^\circ\pm0.052^\circ,\quad
\theta^Q_{13}=0.199^\circ\pm0.008^\circ. 
\end{eqnarray}
The CP violating phase has also been determined with
$\delta^Q=68.9^\circ$~\cite{FERMILAB-PUB-10-665-PPD}.

Considerable experimental data on lepton mixing have also been accumulated including the recent data from DAYA-BAY collaboration\cite{daya}. The
global, $1\sigma(3\sigma)$, fit from neutrino oscillation data pre-DAYA-Bay data
gives~\cite{fogli},
\begin{eqnarray} \left (
\begin{array}{lll}
0.824^{+0.011}_{-0.010}(^{+0.032}_{-0.032})\;\;&0.547^{+0.016}_{-0.014}(^{+0.047}_{-0.044})\;\;&0.145^{+0.022}_{-0.031}(^{+0.065}_{-0.113})\\
0.500^{+0.027}_{-0.021}(^{+0.076}_{-0.071})\;\;&0.582^{+0.050}_{-0.023}(^{+0.139}_{-0.069})\;\;&0.641^{+0.061}_{-0.023}(^{+0.168}_{-0.063})\\
0.267^{+0.044}_{-0.027}(^{+0.123}_{-0.088})\;\;&0.601^{+0.048}_{-0.022}(^{+0.133}_{-0.069})\;\;&0.754^{+0.052}_{-0.020}(^{+0.143}_{-0.054})\\
\end{array} \right)\;.\label{vvv}
\end{eqnarray}
and the mixing angles are given by~\cite{fogli}
\begin{eqnarray}
&&\theta_{12}^L=33.59^\circ\pm1.02^\circ (\pm3.05^\circ),\quad
\theta_{23}^L=40.40^\circ\pm3.17^\circ (\pm8.63^\circ),\nonumber\\
&&\theta_{13}^L=8.33^\circ\pm1.39^\circ (\pm3.57^\circ).
\label{langle}
\end{eqnarray}

The recent measured non-zero $\theta_{13}$ at a 5.2$\sigma$ level by the DAYA-BAY collaboration is\cite{daya} $\sin^2(2\theta_{13}) = 0.092\pm 0.016(stat.) \pm 0.005(syst)$. Translating
into the angle $\theta_{13}$, it is $\theta_{13} = 8.8^\circ\pm 0.8^\circ$. This values agrees with global fit value very well, but with a smaller error bar.
At present there is no experimental data on the CP violating Dirac phase $\delta^L$ and Majorana phases $\alpha_i$.

The mixing angles for quark and lepton sectors, a
priori, are unrelated. If there is a way to connect the seemingly
independent mixing angles in these two sectors, it
would gain more insights about fermion mixing. Indeed there is a
very nice way to make the connection via the so called quark-lepton
complementarity (QLC)~\cite{smirnov,qlc,pheno,model}.
Here the QLC relations refer to
\begin{eqnarray}
\theta_{12}^Q+\theta_{12}^L=\frac{\pi}{4},
\quad\theta_{23}^Q+\theta_{23}^L=\frac{\pi}{4}.\label{qlc}
\end{eqnarray}
The third angles are approximately zero, $\theta_{13}^Q\sim\theta_{13}^L\sim 0$.

The first two relations hold within experimental errors,
\begin{eqnarray}
\theta_{12}^Q+\theta_{12}^L= 46.606^\circ\pm1.019^\circ,
\quad\theta_{23}^Q+\theta_{23}^L= 42.746^\circ\pm3.171^\circ.
\end{eqnarray}

At present the above relations are still at the phenomenological level. It is
far from having a complete theoretical understanding although there are attempt to build theoretical models~\cite{model}. Even at the phenomenological level,
there are some questions to address about these relations. One of them is that there are different ways to parameterize
the mixing matrices. The relation between the angles may only hold in a particular
parametrization~\cite{parametrization-dep}. Therefore these relations seem to be parametrization dependent.
Let us illustrate this point by working out the values of the angles
in the original KM parametrization~\cite{km},
\begin{eqnarray}
V_{KM} = \left ( \begin{array}{ccc} c_1& - s_1 c_3& -s_1 s_3\\s_1c_2&c_1c_2c_3 - s_2s_3 e^{i\delta}&c_1c_2s_3 + s_2c_3 e^{i\delta}\\
s_1s_2&c_1s_2c_3 + c_2 s_3 e^{i\delta}& c_1s_2 s_3 - c_2 c_3 e^{i\delta}\end{array}
\right )\;.
\end{eqnarray}

Using the observed values for the mixing matrices, one would obtain
\begin{eqnarray}
&&\theta^Q_1 = 13.023^\circ\pm0.038^\circ\;,\;\;\theta^Q_2 = 2.192^\circ\pm0.059^\circ\;,\;\;\theta^Q_3 =0.882^\circ\pm0.036^\circ\nonumber\\
&&\theta^L_1 = 34.485^\circ\pm1.028^\circ\;,\;\;\theta^L_2 = 28.086^\circ\pm3.762^\circ\;,\;\;\theta^L_3 =14.830^\circ\pm2.423^\circ\;.
\end{eqnarray}
We see that only  $\theta^Q_1 + \theta^L_1$ is close to $\pi/4$, and the other two angles sums do not have the complementarity relations.

It has been shown that there are nine independent ways to parameterize the mixing matrices~\cite{xing}. We have
checked, in details, whether similar complementarity relations hold in some of the other parameterizations.
We find that only the PDG parametrization
has the complementarity property for two different angles. To avoid the complementarity relations be parametrization dependent,
it would be more meaningful to find the relations using quantities which are reparametrization invariant. We will work the scenario that
there are two complementarity relations discussed above.
To this end we find the magnitudes of the elements in the mixing matrices convenient quantities. They can be used to represent the
complementarity relations.  In the following we derive such relations.

We will take the usual complementarity relations in eq.(\ref{qlc}) as the starting point. By taking sine and cosine on both sides, we obtain
\begin{eqnarray}
&&\sin(\theta^Q_{12} + \theta^L_{12}) = \sin\theta^Q_{12}\cos\theta^L_{12} + \cos\theta^Q_{12}\sin\theta^L_{12} = {1\over \sqrt{2}}\;,\nonumber\\
&&\cos(\theta^Q_{12} + \theta^L_{12}) = \cos\theta^Q_{12}\cos\theta^L_{12} - \sin\theta^Q_{12}\sin\theta^L_{12} = {1\over \sqrt{2}}\;;\nonumber\\
&&\sin(\theta^Q_{23} + \theta^L_{23}) = \sin\theta^Q_{23}\cos\theta^L_{23} + \cos\theta^Q_{23}\sin\theta^L_{23} = {1\over \sqrt{2}}\;,\nonumber\\
&&\cos(\theta^Q_{23} + \theta^L_{23}) = \cos\theta^Q_{23}\cos\theta^L_{23} - \sin\theta^Q_{23}\sin\theta^L_{23} = {1\over \sqrt{2}}\;.
\end{eqnarray}

Using the relations between the angles and elements in the mixing matrices, and taking the ratios of the first two and the last two equations above, we have
\begin{eqnarray}
&&{\tan\theta^Q_{12} + \tan\theta^L_{12}\over 1 -\tan\theta^Q_{12}\tan \theta^L_{12}} = {|V_{us}/V_{ud}| + |V_{e2}/V_{e1}| \over 1 - |V_{us}/V_{ud}||V_{e2}/V_{e1}|} = 1\;,\nonumber\\
&&{\tan\theta^Q_{23} + \tan\theta^L_{23}\over 1 -\tan\theta^Q_{23}\tan \theta^L_{23}} = {|V_{cb}/V_{tb}| + |V_{\mu 3}/V_{\tau 3}| \over 1 - |V_{cb}/V_{tb}||V_{\mu 3}/V_{\tau 3}|} = 1\;.
\end{eqnarray}

The complementarity relations can now be written as
\begin{eqnarray}
&&{|V_{us}|\over |V_{ud}|} + {|V_{e2}|\over |V_{e1}|} + {|V_{us}|\over |V_{ud}|}{|V_{e2}|\over |V_{e1}|} = 1\;,\nonumber\\
&&{|V_{cb}|\over |V_{tb}|} + {|V_{\mu 3}|\over |V_{\tau 3}|}+ {|V_{cb}|\over |V_{tb}|}{|V_{\mu 3}|\over |V_{\tau 3}|} = 1\;.
\end{eqnarray}

The above relations are the new reparametrization invariant complementarity relations. These relations may tell more about the
mixing matrix elements. One can solve these relations to express ratios of elements of lepton mixing matrix in terms of the
ratios of elements of quark mixing matrix to obtain
\begin{eqnarray}
{|V_{e2}|\over |V_{e1}|} = {1 - |V_{us}|/|V_{ud}|\over 1 + |V_{us}|/|V_{ud}|}\;,\;\;{|V_{\mu 3}|\over |V_{\tau 3}|} = {1 - |V_{cb}|/|V_{tb}|\over 1 + |V_{cb}|/|V_{tb}|}\;.\label{key}
\end{eqnarray}
Similarly one can express ratios of quark mixing matrix elements in terms of the lepton mixing matrix elements.

Experimentally, the quark mixing matrix elements are more precisely known, therefore one can take them as input to predict the lepton mixing matrix element. We
have
\begin{eqnarray}
{|V_{e2}|\over |V_{e1}|} = 0.624369\pm0.00095\;,\;\;{|V_{\mu
3}|\over |V_{\tau 3}|} = 0.921165\pm0.00166\;.\label{number}
\end{eqnarray}
Notice that $|V_{\tau 3}| > |V_{\mu 3}|$. This is consistent with data from neutrino oscillation at 2$\sigma$ and 1$\sigma$ level for the first and the second ratios in the above equations, respectively.

The tribimaximal mixing is a good approximation for the lepton
mixing matrix~\cite{tri}. Data from global fit and the recent
DAYA-BAY measurement show that the tribimaximal mixing pattern, which predicts $\theta_{13} = 0$, has
to be modified. There are models prefer that the elements in the
second column to be all  equal to $1/\sqrt{3}$~\cite{tri-vary}. If
this is true then combining $V_{e2} = 1/\sqrt{3}$ and
eq.(\ref{number}), one would obtain $|V_{e1}|^2+|V_{e2}|^2>1$. This
indicates that the QLC is not consistent with this type of models.
There are also speculations that other columns or row in the
tribimaximal mixing is kept unaltered but other elements are
modified~\cite{tri-vary}. If one keeps the first column of the
tribimaximal mixing matrix unaltered, one would have $V_{e1} =
2/\sqrt{6}$. Coming with eq.(\ref{number}), one would obtain $V_{e3}
= 0.27$. This predicts too large a $V_{e3}$ outside the $1\sigma$
region allowed by the pre-DAYA-BAY global data fit and more
from present DAYA-BAY data. Also if one keeps the second row of
tribimaximal mixing unaltered, one would then predict $|V_{\mu 3}|^2
+ |V_{\tau 3}|^2>1$ which is not allowed. If one keeps the third row
of tribimaximal mixing unaltered~\cite{lee}, then one would obtain
also too large a $V_{e3} = 0.275$. Satisfaction of QLC requirement
would require modification of the tribimaximal mixing
significantly~\cite{deviation}.

In the above relations, the elements $V_{e3}$ and $V_{ub}$ do not show up directly. To have some information about the 13 entries of the mixing matrices, let us consider the
unitarity of the first row and the third column. Eq.(\ref{key}) suggest that one can write
\begin{eqnarray}
&&|V_{e1}| = {a\over \sqrt{2}}(|V_{ud}| + |V_{us}|)\;,\;\;|V_{e2}| = {a\over \sqrt{2}}(|V_{ud}| - |V_{us}|)\;,\nonumber\\
&&|V_{\tau 3}| = {b\over \sqrt{2}}(|V_{tb}| + |V_{cb}|)\;,\;\;|V_{\mu 3}| = {b\over \sqrt{2}}(|V_{tb}| - |V_{cb}|)\;.
\end{eqnarray}
One then obtains
\begin{eqnarray}
\sum_i|V_{ei}|^2 = a^2(|V_{ud}|^2 + |V_{us}|^2) + |V_{e3}|^2 = b^2(|V_{cb}|^2 + |V_{tb}|^2) + |V_{e3}|^2 =1\;.
\end{eqnarray}
$a$ and $b$ are related by $b^2 = a^2(|V_{ud}|^2 +
|V_{us}|^2)/(|V_{cb}|^2 + |V_{tb}|^2) = (1.0000\pm0.0004)a^2$.

The size for $V_{e3}$ depends on what $a$ in the form: $|V_{e3}| =
\sqrt{1 - a^2(|V_{ud}|^2 + |V_{us}|^2)}$. Using current $1\sigma$
allowed value for $|V_{e3}|$, a is determined to be smaller than
0.99.

There may be modifications to the complementarity relations. The modifications can be written as $\theta^Q + \theta^L = \pi/4 + \alpha$. In this case the reparametrization invariant
relations will be modified to
\begin{eqnarray}
&&{|V_{us}|\over |V_{ud}|} + {|V_{e2}|\over |V_{e1}|} +
{|V_{us}|\over |V_{ud}|}{|V_{e2}|\over |V_{e1}|}
= 1 + \left (1 - {|V_{us}|\over |V_{ud}|}{|V_{e2}|\over |V_{e1}|}\right ) {\tan\alpha_{12}\over 1 - \tan\alpha_{12}}\;,\nonumber\\
&&{|V_{cb}|\over |V_{tb}|} + {|V_{\mu 3}|\over|V_{\tau 3}|}+ {|V_{cb}|\over|V_{tb}|}{|V_{\mu 3}|\over |V_{\tau 3}|}
= 1 + \left (1 - {|V_{cb}|\over |V_{tb}|}{|V_{\mu 3}|\over |V_{\tau 3}|}\right ) {\tan\alpha_{23}\over 1 - \tan\alpha_{23}}\;.
\end{eqnarray}

\begin{figure}[htbp]
\centering
\includegraphics[width = 0.6\textwidth]{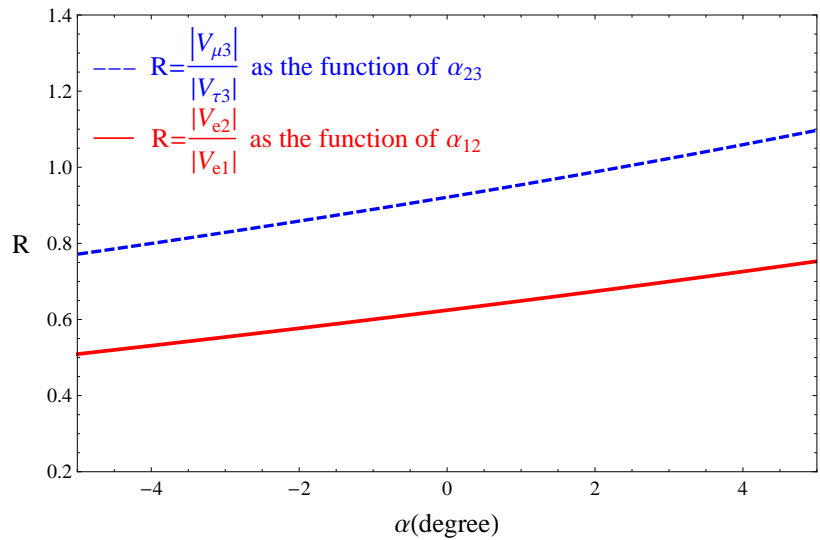}\hspace{0.5cm}
\caption{Ratios of $|V_{e2}|/|V_{e1}|$ and $|V_{\mu3}|/|V_{\tau3}|$ as functions of deviations $\alpha_{12}$ and $\alpha_{23}$, respectively.}
\end{figure}

Expressing lepton mixing matrix elements in terms of the quark mixing elements and the deviations, we have
\begin{eqnarray}
&&{|V_{e2}|\over |V_{e1}|} = {1 + \tan\alpha_{12} - |V_{us}|/|V_{ud}|(1- \tan\alpha_{12}) \over 1 - \tan\alpha_{12} + |V_{us}|/|V_{ud}|(1 + \tan\alpha_{12})}\;,\nonumber\\
&&{|V_{\mu 3}|\over |V_{\tau 3}|} = {1 + \tan\alpha_{23} -
|V_{cb}|/|V_{tb}|(1- \tan\alpha_{23}) \over 1 - \tan\alpha_{23} +
|V_{cb}|/|V_{tb}|(1 + \tan\alpha_{23})}\;.
\end{eqnarray}

In Fig. 1 we show how $|V_{e2}|/|V_{e1}|$ and $|V_{\mu 3}|/|V_{\tau 3}|$ depend on $\alpha_{12}$ and $\alpha_{23}$. We see that a small deviation away from the
complementarity relation can cause sizeable difference in the predicted neutrino mixing matrix elements. With more precisely measured mixing matrix elements in both the quark and lepton sectors, the QLC and deviations can be studied more which may give some hints on theoretical models for quark and lepton mixing matrices.

This work was supported in part by NSC and NCTS of ROC, NNSF and SJTU 985 grants of PRC.

\end{document}